\documentclass[aps,prl,twocolumn,showpacs,floatfix,longbibliography]{revtex4-1}
\usepackage{graphicx,amsfonts,amssymb,amsmath}
\usepackage{textcomp}
\usepackage{esint}
\usepackage[colorlinks,linktocpage,bookmarks=false,citecolor=blue,linkcolor=red,urlcolor=blue]{hyperref}
\usepackage[T1]{fontenc}
\usepackage[dvipsnames]{xcolor}
\usepackage[english]{babel}
\usepackage{afterpage}
\usepackage{empheq}

\allowdisplaybreaks

\def\kin{k_{\rm in}} 
\def\Msol{M_{\rm sol}} 
\def\Mcl{M_{\rm cl}} 
\def\epsabs{\eps_{\rm abs}}
\def\epsrel{\eps_{\rm rel}}

\begin{document}


\def\rhoeq{\hat\rho_{\rm eq}}

\newcommand{\marge}[1]{\marginpar{\scriptsize #1}}
\newcommand{\remarque}[1]{\marginpar{\scriptsize Remarque}{\it [#1]}}
\newcommand{\new}[1]{{\bf #1}}
\newlength{\textlarg}
\newcommand{\redbar}[1]{\textcolor{red}{\st{#1}}} 
\newcommand{\bluebar}[1]{\textcolor{blue}{\st{#1}}} 

\newcommand{\beq}{\begin{equation}}
\newcommand{\eeq}{\end{equation}}
\newcommand{\bleq}{\begin{eqnarray}}
\newcommand{\eleq}{\end{eqnarray}} 
\newcommand{\bfig}{\begin{figure}}
\newcommand{\efig}{\end{figure}}
\newcommand{\bline}{\begin{multline}}
\newcommand{\eline}{\end{multline}}
\newcommand{\bremark}{\begin{quotation} \noindent \small }
\newcommand{\eremark}{\end{quotation}}
\newcommand{\llbrace}{\left\lbrace}
\newcommand{\rrbrace}{\right\rbrace}
\newcommand{\lbraket}{\left[}
\newcommand{\rbraket}{\right]}
\newcommand{\llangle}{\left\langle}
\newcommand{\rrangle}{\right\rangle} 

\newcommand{\Tr}{{\rm Tr}} 
\newcommand{\tr}{{\rm tr}} 
\newcommand{\sgn}{\,{\rm sgn}} 
\newcommand{\mean}[1]{\langle #1 \rangle}
\newcommand{\commu}[2]{[#1,#2]} 
\newcommand{\bra}[1]{\langle#1|}
\newcommand{\ket}[1]{|#1\rangle}
\newcommand{\braket}[2]{\langle #1|#2\rangle}
\newcommand{\dbraket}[3]{\langle #1|#2|#3\rangle}
\newcommand{\tens}[1]{\overleftrightarrow{#1}}  
\newcommand{\vac}{|{\rm vac}\rangle} 
\newcommand{\bravac}{\langle{\rm vac}|}
\newcommand{\const}{{\rm const}} 
\newcommand{\atanh}{\,{\rm atanh}}
\newcommand{\cotanh}{\,{\rm cotanh}}

\newcommand{\ie}{i.e.\xspace}
\newcommand{\iet}{i.e.}
\newcommand{\eg}{e.g.\xspace}
\newcommand{\cc}{{\rm c.c.}} 
\newcommand{\hc}{{\rm h.c.}} 
\newcommand{\etal}{{\it et al. }}
\newcommand\eme{$^{\mbox{\footnotesize ème}}$\xspace}

\newcommand{\jhatbf}{\hat {\textbf \jold}} 
\newcommand{\Jhatbf}{\hat {\textbf \J}} 
\newcommand{\jhat}{\hat {\jmath}} 
\newcommand{\Jhat}{\hat {J}} 
\newcommand{\jbf}{\textbf j}
\newcommand{\Jbf}{\textbf J}

\def\chibf{\boldsymbol{\chi}}
\def\down{\downarrow}
\def\eps{\epsilon}
\def\gam{\gamma} 
\def\alphabf{\boldsymbol{\alpha}}
\def\phibf{\boldsymbol{\phi}}
\def\varphibf{\boldsymbol{\varphi}}
\def\varphibfs{\boldsymbol{\varphi}_<}
\def\varphibfl{\boldsymbol{\varphi}_>}
\def\varphis{\varphi_{<}}
\def\varphil{\varphi_{>}}
\def\psibf{\boldsymbol{\psi}}
\def\thetabf{\boldsymbol{\theta}}
\def\Ome{\Omega}
\def\omeD{{\omega_D}} 
\def\bfOme{\boldsymbol{\Omega}} 
\def\Omebf{\boldsymbol{\Omega}} 
\def\lamb{\lambda}
\def\Lamb{\Lambda}
\def\sig{\sigma}
\def\Sig{\Sigma}
\def\sigp{{\sigma'}} 
\def\bfsig{\boldsymbol{\sigma}} 
\def\sigbf{\boldsymbol{\sigma}} 
\def\bfSig{\boldsymbol{\Sigma}} 
\def\The{\Theta} 
\def\up{\uparrow}

\def\epsk{\epsilon_{\bf k}} 
\def\xik{\xi_{\bf k}} 
\def\txik{\tilde\xi_{\bf k}} 
\def\xip{\xi_{\bf p}} 
\def\xiq{\xi_{\bf q}} 
\def\xikq{\xi_{{\bf k}+{\bf q}}} 
\def\Ek{E_{\bf k}} 
\def\Ep{E_{\bf p}}
\def\Eq{E_{\bf q}}
\def\Heff{\hat H_{\rm eff}}
\def\Hem{\hat H_{\rm em}}
\def\Hint{\hat H_{\rm int}}
\def\Hloc{\hat H_{\rm loc}}
\def\HMF{\hat H_{\rm MF}}
\def\Sem{S_{\rm em}}
\def\SMF{S_{\rm MF}} 
\def\SHF{S_{\rm HF}} 
\def\SRPA{S_{\rm RPA}} 
\def\Sint{S_{\rm int}} 
\def\Sloc{S_{\rm loc}}
\def\TN{T_{\rm N}} 
\def\TNHF{T^{\rm HF}_{\rm N}} 
\def\Zloc{Z_{\rm loc}} 
\def\ZMF{Z_{\rm MF}} 
\def\ZHF{Z_{\rm HF}} 
\def\ZRPA{Z_{\rm RPA}} 
\def\RPA{{\rm RPA}}
\def\loc{{\rm loc}} 
\def\pp{{\rm pp}}
\def\ph{{\rm ph}} 
\def\ch{{\rm ch}}
\def\sp{{\rm sp}} 
\def\qtf{q_{\rm TF}}
\def\epstf{\eps^{}_{\rm TF}} 
\def\epsrpa{\eps^{}_{\rm RPA}} 
\def\chinnzpp{\chi_{nn}^{0}{}\!\!\!''}

\def\half{\frac{1}{2}}
\def\dhalf{\dfrac{1}{2}}
\def\third{\frac{1}{3}} 
\def\quarter{\frac{1}{4}}

\def\qr{{\bf q}\cdot{\bf r}}
\def\wt{\omega t} 

\def\a{{\bf a}}
\def\b{{\bf b}}
\newcommand{\cv}{{\bf c}} 
\def\e{{\bf e}}
\def\f{{\bf f}}
\def\g{{\bf g}}
\def\h{{\bf h}}
\def\jold{\char"11}
\def\j{{\bf j}}
\def\k{{\bf k}}
\def\l{{\bf l}}
\def\m{{\bf m}}
\def\n{{\bf n}} 
\def\p{{\bf p}} 
\def\q{{\bf q}}
\def\r{{\bf r}}
\def\t{{\bf t}}
\def\u{{\bf u}}
\newcommand{\vv}{{\bf v}}
\def\x{{\bf x}}
\def\y{{\bf y}} 
\def\z{{\bf z}} 
\def\A{{\bf A}}
\def\B{{\bf B}}
\def\D{{\bf D}} 
\def\E{{\bf E}} 
\def\F{{\bf F}} 
\def\H{{\bf H}}  
\def\J{{\bf J}}
\def\K{{\bf K}} 

\def\G{{\bf G}}
\def\L{{\bf L}}
\def\M{{\bf M}}  
\def\O{{\bf O}} 
\def\P{{\bf P}} 
\def\Q{{\bf Q}} 
\def\R{{\bf R}}
\def\S{{\bf S}}
\def\U{{\bf U}} 
\def\V{{\bf V}} 
\def\X{{\bf X}} 
\def\Y{{\bf Y}} 
\def\epsbf{\boldsymbol{\epsilon}}
\def\betabf{\boldsymbol{\beta}}
\def\mubf{\boldsymbol{\mu}}
\def\nablabf{\boldsymbol{\nabla}}
\def\rhobf{\boldsymbol{\rho}}
\def\sigmabf{\boldsymbol{\sigma}} 
\def\Pibf{\boldsymbol{\Pi}}
\def\pibf{\boldsymbol{\pi}}

\def\para{\parallel}
\def\kpara{{k_\parallel}}
\def\kperp{{k_\perp}} 
\def\kperpp{{k_\perp'}} 
\def\qperp{{q_\perp}} 
\def\tperp{{t_\perp}} 

\def\w{\omega}
\def\wn{\omega_n}
\def\wm{\omega_m}
\def\wnu{\omega_\nu}
\def\wp{\omega_p} 
\def\dmu{{\partial_\mu}}
\def\dnu{{\partial_\nu}}
\def\dl{{\partial_l}}  
\def\dt{\partial_t} 
\def\tdt{\tilde\partial_t}
\def\dk{\partial_k}
\def\tdk{\tilde\partial_k}
\def\dx{\partial_x}
\def\dy{\partial_y} 
\def\dtau{{\partial_\tau}}  
\def\det{{\rm det}} 
\def\Pf{{\rm Pf}}
\def\diag{{\rm diag}}

\def\dsum{\displaystyle \sum}
\def\dint{\displaystyle \int} 
\def\intt{\int_{-\infty}^\infty dt} 
\def\inttp{\int_{-\infty}^\infty dt'} 
\def\intk{\int_{\bf k}} 
\def\intkd{\int \frac{d^dk}{(2\pi)^d}}
\def\intq{\int_{\bf q}} 
\def\intr{\int d^dr}  
\def\dintr{\displaystyle \int d^dr} 
\def\intrp{\int d^dr'}
\def\dinttau{\displaystyle \int_0^\beta d\tau}
\def\dinttaup{\displaystyle \int_0^\beta d\tau'}
\def\inttau{\int_0^\beta d\tau}
\def\inttaup{\int_0^\beta d\tau'}
\def\intx{\int d^{d+1}x} 
\def\inttaur{\int_0^\beta d\tau \int d^dr}
\def\intinf{\int_{-\infty}^\infty}
\def\dinttaur{\displaystyle \int_0^\beta d\tau \int d^dr}
\def\dintinf{\displaystyle \int_{-\infty}^\infty}
\def\intw{\int_{-\infty}^\infty \frac{d\w}{2\pi}}
\def\sumr{\sum_{\bf r}} 

\def\calA{{\cal A}}
\def\calAbf{\bm{{\cal A}}}
\def\calB{{\cal B}} 
\def\calC{{\cal C}} 
\def\dt{\partial_t}
\def\calD{{\cal D}}
\def\calE{{\cal E}}
\def\calF{{\cal F}} 
\def\calFbf{\bm{{\cal F}}}
\def\calG{{\cal G}}
\def\calH{{\cal H}}
\def\calI{{\cal I}}
\def\calJ{{\cal J}}
\def\calK{{\cal K}}
\def\calL{{\cal L}} 
\def\calM{{\cal M}} 
\def\calN{{\cal N}}
\def\calO{{\cal O}}
\def\calP{{\cal P}}  
\def\calR{{\cal R}} 
\def\calS{{\cal S}}
\def\calT{{\cal T}}
\def\calU{{\cal U}}
\def\calV{{\cal V}}
\def\calX{{\cal X}} 
\def\calY{{\cal Y}} 
\def\calZ{{\cal Z}} 

\def\calbfB{{\bf \cal B}}
\def\calbfF{{\bf \cal F}}

\def\tT{{\tilde T}}
\def\talpha{{\tilde\alpha}}
\def\tdelta{{\tilde\delta}}
\def\teta{{\tilde\eta}} 
\def\tlamb{{\tilde\lambda}}
\def\tmu{{\tilde\mu}}
\def\tphibf{{\tilde\phibf}}
\def\trho{{\tilde\rho}}
\def\tvarphibf{{\tilde\varphibf}} 
\def\tw{{\tilde\omega}}
\def\twn{{\tilde\omega_n}}
\def\twnu{{\tilde\omega_\nu}}

\def\asinh{{\rm asinh}} 

\graphicspath{{./figures/}}

\title{Nonperturbative functional renormalization-group approach to the sine-Gordon model and the Lukyanov-Zamolodchikov conjecture}

\author{R. Daviet and N. Dupuis}
\affiliation{Sorbonne Universit\'e, CNRS, Laboratoire de Physique Th\'eorique de la Mati\`ere Condens\'ee, LPTMC, F-75005 Paris, France}

\date{December 5, 2018} 

\begin{abstract}
We study the quantum sine-Gordon model within a nonperturbative functional renormalization-group approach (FRG). This approach is benchmarked by comparing our findings for the soliton and lightest breather (soliton-antisoliton bound state) masses to exact results. We then examine the validity of the Lukyanov-Zamolodchikov conjecture for the expectation value $\mean{e^{\frac{i}{2}n\beta\varphi}}$ of the exponential fields in the massive phase ($n$ is integer and $2\pi/\beta$ denotes the periodicity of the potential in the sine-Gordon model). We find that the minimum of the relative and absolute disagreements between the FRG results and the conjecture is smaller than 0.01.
\end{abstract}

\maketitle

\paragraph{Introduction.} The quantum sine-Gordon model~\cite{[{For an introduction to the quantum sine-Gordon model see }]Rajaraman_book} describes many physical systems. In condensed matter it is widely used to understand the phase diagram and the low-energy properties of one-dimensional quantum fluids~\cite{Gogolin_book,Tsvelik_book,Giamarchi_book} and has applications that range from strongly correlated electron systems to cold atoms. In high-energy physics it is related to the massive Thirring model describing Dirac fermions with a self interaction~\cite{Coleman75}. The sine-Gordon model can also be viewed as a two-dimensional model of classical statistical mechanics. In particular it describes the Berezinskii-Kosterlitz-Thouless (BKT) transition which occurs in the XY spin model and more generally in two-dimensional systems with a two-component order parameter with an O(2) symmetry~\cite{Berezinskii71,*Berezinskii72,Kosterlitz73,Kosterlitz74}. 

The Hamiltonian of the quantum sine-Gordon model is defined by 
\beq
\hat H = \int dx \biggl\lbrace \half \hat\Pi^2 + \half \left(\frac{\partial\hat\varphi}{\partial x}\right)^2 - u \cos(\beta\hat\varphi) \biggr\rbrace , 
\label{ham} 
\eeq 
where $\hat\Pi$ and $\hat\varphi$ satisfy canonical commutation relations, $[\hat\varphi(x),\hat\Pi(x')]=i\delta(x-x')$. Regularization with a UV momentum cutoff $\Lambda$ is implied and $u/\Lamb^2,\beta>0$ are dimensionless parameters. The phase diagram consists of a gapless phase with massless (anti)soliton excitations for $\beta^2\geq 8\pi$ (and $u\to 0$) and a gapped phase with massive (anti)soliton excitations for $\beta^2\leq 8\pi$. The soliton and the antisoliton carry the topological charge $Q=1$ and $-1$, respectively~\cite{not5}. They attract for $\beta^2\leq 4\pi$ and can form bound states, called breathers, with topological charge $Q=0$.
The phase transition between the two phases is of BKT type~\cite{Berezinskii71,*Berezinskii72,Kosterlitz73,Kosterlitz74}. 

The sine-Gordon model is one of the most studied integrable models; its spectrum, thermodynamics and scattering properties are well understood~\cite{Faddeev78,Sklyanin80,Zamolodchikov77,Zamolodchikov79,Smirnov_book}. However not everything is known and many quantities can be obtained only from nonexact (e.g. perturbative) methods~\cite{Gogolin_book,Tsvelik_book,Giamarchi_book}. In particular in the massive phase the amplitude of the fluctuations about the mean value $\mean{\hat\varphi}=0$ is not known exactly. It has been conjectured by Lukyanov and Zamolodchikov that~\cite{Lukyanov97} 
\begin{widetext} 
\beq 
\mean{e^{i\sqrt{8\pi}a\hat\varphi}} = \left[ \frac{ \Gamma(1-K) }{ \Gamma(K) } \frac{\pi u}{2(b\Lamb)^2} \right]^{\frac{a^2}{1-K}} 
\exp \biggl\lbrace \int_0^\infty \frac{dt}{t} \biggl[ \frac{\sinh^2(2a\sqrt{K}t)}{2\sinh(Kt) \sinh(t) \cosh[(1-K)t]} - 2a^2 e^{-2t} \biggr] \biggr\rbrace ,
\label{conjecture}
\eeq
\end{widetext} 
where $|\Re(a)|<1/2\sqrt{K}$ and $K=\beta^2/8\pi$ is the ``Luttinger parameter''~\cite{Giamarchi_book} (the massive phase corresponds to $K<1$). Equation~(\ref{conjecture}) is exact for $a=\sqrt{K}$, $K=1/2$ and in the semiclassical limit $K\to 0$~\cite{not4}. Additional arguments supporting the conjecture were  presented in~\cite{Fateev97,Fateev98}. From the equivalence between the sine-Gordon model and the massive Thirring model Eq.~(\ref{conjecture}) was shown to be correct to first order in $u$~\cite{Poghossian00,Mkhitaryan00}. Further evidence of the correctness of~(\ref{conjecture}), in particular for not too large values of $a$, was provided by a numerical study in a finite volume~\cite{Bajnok00} and variational perturbation theory~\cite{Lu04}. 

In this Letter, we examine the validity of the Lukyanov-Zamolodchikov conjecture using a nonperturbative functional renormalization-group approach (FRG)~\cite{not6,[{Nonperturbative flow equations have been obtained from continuous unitary transformations but the determination of the the spectrum (soliton and breather's masses) and the validity of the Lukyanov-Zamolodchikov conjecture have not been addressed. See }] Kehrein99,*Kehrein01}.
We go beyond previous FRG approaches~\cite{Nagy09,Pangon12,Pangon11,Bacso15} and, in order to benchmark our approach, first  compute the mass $\Msol$ of the (anti)soliton as well as that ($M_1$) of the lightest breather. We then turn to the computation of the expectation value $\mean{e^{\frac{i}{2}n\beta\varphi}}=\mean{e^{in\sqrt{2\pi K}\varphi}}$ ($n$ integer) of the exponential fields. We confirm the Lukyanov-Zamolodchikov conjecture with an accuracy, defined as the minimum of the relative and absolute disagreements between the FRG results and the conjecture, of 0.01.

\paragraph{FRG approach.} From now on we adopt the point of view of classical statistical mechanics (or Euclidean field theory) where the sine-Gordon model is defined by the partition function 
\beq
\calZ[J] = \int\calD[\varphi] \, e^ { -\int d^2 r \bigl\lbrace \half (\nablabf \varphi)^2 - u \cos(\beta\varphi) - J \varphi \bigr\rbrace } ,
\label{ZJ}
\eeq
with $\varphi(\r)$ a classical field and $\r$ a two-dimensional coordinate. 
$J$ is an external source allowing us to obtain the expectation value $\phi(\r)=\mean{\varphi(\r)}=\delta\ln \calZ[J]/\delta J(\r)$ by functional derivation. Most physical quantities can be obtained from the free energy $-\ln \calZ[J]$ or, equivalently, the effective action (or Gibbs free energy) 
\beq
\Gamma[\phi] = - \ln \calZ[J] + \int d^2r\, J\phi 
\eeq
defined as the Legendre transform of $\ln\calZ[J]$. 

\begin{figure}[b]
\centerline{\includegraphics[width=7cm]{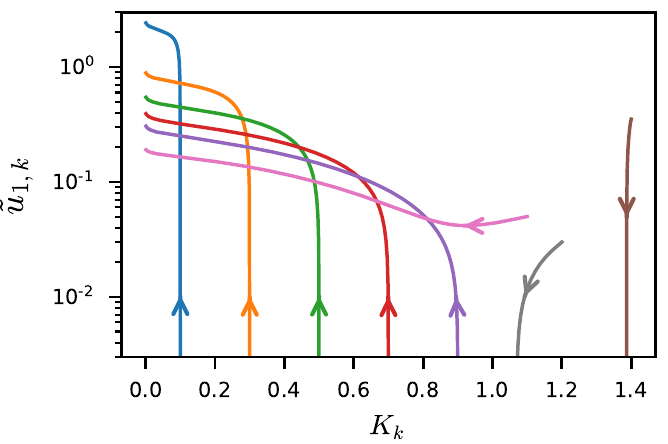}}
\caption{Flow diagram of the sine-Gordon model projected onto the plane $(K_k,\tilde u_{1,k})$ where $\tilde u_{1,k}$ is the first harmonic of the potential $\tilde U_k(\phi)$ and $K_k$ the running Luttinger parameter. There is an attractive line of fixed points for $\tilde u_{1,k}=0$ and $K_k>1$ which terminates at the BKT point $(\tilde u_{1,k}=0,K_k=1$).}
\label{fig_flow_diagram} 
\end{figure}

\begin{figure}[t]
\centerline{\includegraphics[width=4.2cm]{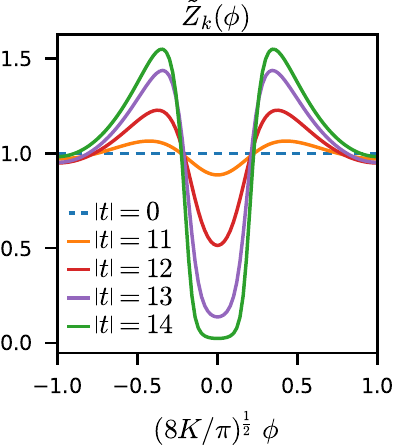}
\includegraphics[width=4.2cm]{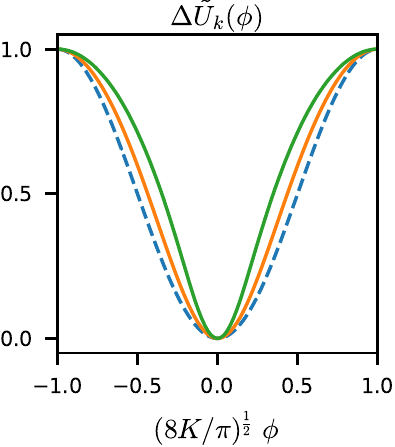}}
\caption{$\tilde Z_k(\phi)$ and $\tilde U_k(\phi)$ for various values of $t=\ln(k/\kin)$. $\Delta\tilde U_k(\phi)$ is given by $\tilde U_k(\phi)-\tilde U_k(0)$ normalized so that $\Delta\tilde U_k(\pm\sqrt{\pi/8K})=1$. In Figs.~\ref{fig_ZU} and \ref{fig_mass_K}, $\Lamb=1$ and $u/\Lamb^2=10^{-3}$.} 
\label{fig_ZU} 
\end{figure}

We compute $\Gamma[\phi]$ using a Wilsonian nonperturbative FRG approach where fluctuation modes are progressively integrated out in the functional integral~(\ref{ZJ}). This defines a scale-dependent effective action $\Gamma_k[\phi]$ which incorporates fluctuations with momenta between a (running) momentum scale $k$ and the UV scale $\kin\gg\Lamb$. The latter condition implies that the initial value $\Gamma_{\kin}[\phi]=S[\phi]$ coincides, as in mean-field theory, with the microscopic action defined by~(\ref{ZJ}). The effective action of the sine-Gordon model, $\Gamma_{k=0}[\phi]$, is obtained when all fluctuations have been integrated out. The scale-dependent effective action satisfies an exact flow equation which cannot be solved exactly~\cite{[{For reviews on the nonperturbative functional renormalization group, see }]Berges02,*Delamotte12,*Kopietz_book}. A common approximation scheme is the derivative expansion where 
\beq
\Gamma_k[\phi] = \int d^2r \llbrace \half Z_k(\phi) (\nablabf\phi)^2 + U_k(\phi) \rrbrace 
\eeq
is truncated to second order in derivatives. This leads to coupled flow equations for the functions $Z_k(\phi)$ and $U_k(\phi)$, with initial conditions $Z_{\kin}(\phi)=1$ and $U_{\kin}(\phi)=-u\cos(\beta\phi)$, which can be solved numerically. We refer to the Supplemental Material for more detail about the implementation of the FRG approach~\cite{not2}.

It is convenient to consider the dimensionless functions 
\beq
\tilde Z_k(\phi)= \frac{Z_k(\phi)}{Z_k}, \quad \tilde U_k(\phi) = \frac{U_k(\phi)}{Z_k k^2} , 
\eeq 
where $Z_k=\mean{Z_k(\phi)}_\phi$ is obtained by averaging $Z_k(\phi)$ on $]-\beta/\pi,\beta/\pi]$~\cite{not7}. The flow diagram, projected onto the plane $(K_k,\tilde u_{1,k})$ is shown in Fig.~\ref{fig_flow_diagram}. Here $\tilde u_{1,k}$ is the first harmonic of the potential $\tilde U_k(\phi)=-\sum_{n=0}^\infty \tilde u_{n,k} \cos(n\beta\phi)$ and $K_k=K/Z_k$ can be interpreted as a ``running'' Luttinger parameter. In the massive phase, the flow runs into fixed points characterized by functions $\tilde Z^*(\phi)$ and $\tilde U^*(\phi)$ which depend on the parameters $u$ and $K$ (Fig.~\ref{fig_ZU}). While $\tilde U^*(\phi)$ slightly deviates from the cosine form of the initial potential $\tilde U_{\kin}(\phi)=-(u/\kin^2)\cos(\beta\phi)$, we see that $\tilde Z^*(\phi)$ acquires a strong dependence on $\phi$. $Z_k$ diverges as $k^{-2}$ and the running Luttinger parameter $K_k\sim k^2$ vanishes for $k\to 0$.

\paragraph{Benchmarking: soliton and breather masses.}

The smallest excitation gap $M$ of the quantum sine-Gordon model corresponds to the inverse correlation length of the two-dimensional classical model~(\ref{ZJ}), 
\beq
M^2 = \lim_{k\to 0} \frac{U_k''(0)}{Z_k(0)} = \lim_{k\to 0}  k^2 \frac{\tilde U_k''(0)}{\tilde Z_k(0)} .
\eeq 
Since $\tilde U''_k(0)$ converges to a finite value --this property is preserved even if one retains only the first harmonics of $\tilde U_k(\phi)$-- $\tilde Z_k(0)$ must vanish as $k^2$ for $M$ to take a nonzero value in the massive phase. $\tilde Z_k(\phi)$ being a normalized function, $\mean{\tilde Z_k(\phi)}_\phi=1$, this is possible only if $\tilde Z_k(\phi)$ strongly varies with $\phi$. Thus only a functional approach where the coefficient of $(\nablabf\phi)^2$ in the effective action is a function $Z_k(\phi)$, and not a mere $\phi$-independent number, can predict the mass of the lowest excitation. Numerically we observe a rapid convergence of $k^2 \tilde U_k''(0)/\tilde Z_k(0)$ when $k\to 0$ in agreement with a previous study~\cite{Pangon11}.  

\begin{figure}[t]
\centerline{\includegraphics[width=7.5cm]{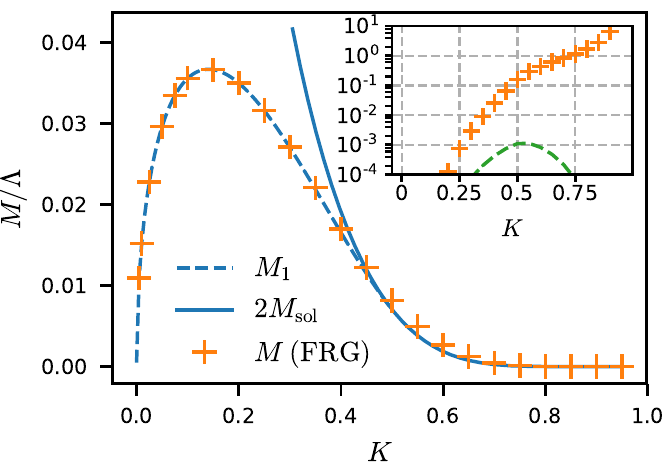}}
\caption{Mass $M$ of the lowest excitation as obtained from the FRG approach. The solid and dashed lines show the exact values of $2\Msol$ and $M_1$ (the latter being defined only for $K\leq 1/2$) [Eqs.~(\ref{Msolexact},\ref{M1exact})]. The inset shows the relative (crosses) and absolute (dashed line) errors of the FRG result.} 
\label{fig_mass_K}
\end{figure}

Only excitations that are in the same topological sector as the ground state, namely $Q=0$, contribute to the mass $M$~\cite{Rajaraman_book}. The lowest excitation in this sector is a soliton-antisoliton pair with mass 
\begin{align}
2\Msol &= b\Lamb \frac{4 \Gamma\left( \frac{K}{2-2K} \right)}{ \sqrt{\pi}\Gamma\left( \frac{1}{2-2K} \right) }
\left[ \frac{ \Gamma(1-K) }{ \Gamma(K) } \frac{\pi u}{2(b\Lamb)^2} \right]^{\frac{1}{2-2K}} , 
\label{Msolexact} 
\end{align} 
when $1/2\leq K\leq 1$ ($\Msol$ is the mass of a single (anti)soliton) and a breather with mass 
\begin{align}
 M_1 &= 2\Msol \sin\left( \frac{\pi}{2} \frac{K}{1-K} \right) ,
\label{M1exact} 
\end{align}
when $0\leq K<1/2$~\cite{Zamolodchikov95}. Here $b$ is a scale parameter which depends on the precise implementation of the UV cutoff $\Lamb$ in Eq.~(\ref{ham}). 
Figure~\ref{fig_mass_K} shows the value of $M$ obtained from FRG (we refer to the Supplemental Material for a discussion of the implementation of the UV cutoff $\Lamb$ and the determination of the scale factor $b$). For $0\leq K\leq 0.4$ our result for the breather mass $M\equiv M_1$ deviates from the exact value by at most 2\%. The agreement becomes nearly perfect for $K\ll 0.4$, which is due to the fact that the initial value $\Gamma^{(2)}_{\kin}(\q,\phi=0)=\q^2+8\pi Ku$ gives the exact breather mass $M_{1,\rm cl}=\sqrt{8\pi Ku}$ in the semiclassical limit $K\to 0$~\cite{not2}. For $0.4\leq K\leq 1$ the agreement between $M$ and the exact value $2\Msol$ is not as good and varies from $\sim 2$\% for $K\simeq 0.4$ to more than 100\% for $K$ near 1. Note however that $M$ goes to zero when $K\to 1$ and the absolute error remains below $10^{-3}$ for all values of $K$ (see the inset in Fig.~\ref{fig_mass_K}). In the immediate vicinity of $K=1$, the behavior of the mass $M$ differs from $2\Msol$ [Eq.~(\ref{Msolexact})] and one recovers the standard BKT scaling characterized by an essential singularity of the correlation length~\cite{Kosterlitz74}.

\paragraph{The Lukyanov-Zamolodchikov conjecture.} 

\begin{figure}
\centerline{\includegraphics[width=7.5cm]{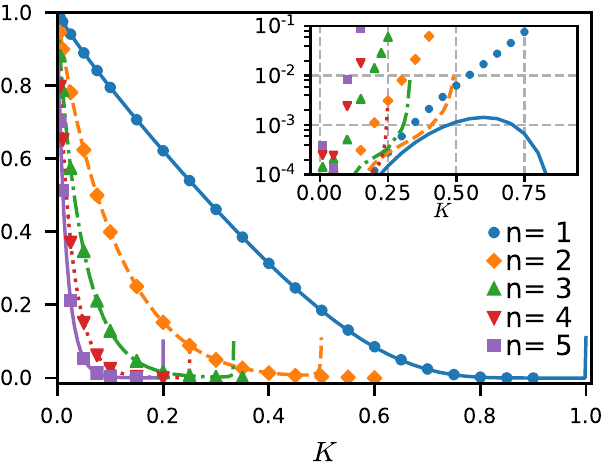}} 
\caption{Expectation value $\mean{e^{in\sqrt{2\pi K}\varphi}}$ as obtained from FRG (symbols) vs the Lukyanov-Zamolodchikov conjecture~(\ref{conjecture}) valid for $K<1/n$ (lines). The inset shows the relative (symbols) and absolute (lines) disagreements between the FRG results and the conjecture, respectively $\epsrel$ and $\epsabs$. ($\Lamb=1$ and $u/\Lamb^2=10^{-4}$.)}
\label{fig_conjecture} 
\end{figure}

To obtain the expectation value of the exponential fields, we consider the partition function~(\ref{ZJ}) in the presence of an external source term $\int d^2 r(h^* e^{i\sqrt{8\pi}a\varphi}+\cc)$ so that $\mean{e^{i\sqrt{8\pi}a\varphi(\r)}}$ can be obtained from $\ln\calZ_k[J,h^*,h]$ by functional differentiation wrt $h^*(\r)$. To second order of the derivative expansion the effective action now reads
\begin{align}
\Gamma_k[\phi,h^*,h] ={}& \int d^2r \biggl\lbrace \half Z_k(\phi,h^*,h) (\nablabf\phi)^2 \nonumber \\ & 
+ U_k(\phi,h^*,h) \biggr\rbrace 
\end{align} 
and
\beq
\mean{e^{i\sqrt{8\pi}a\varphi(\r)}} = - \frac{\partial U_k(\phi=0,h^*,h)}{\partial h^*} \biggl|_{h^*=h=0} . 
\eeq 
From the flow equation of $\Gamma_k[\phi,h^*,h]$ we obtain two coupled equations for $U_k^{(10)}(\phi)\equiv\partial_{h^*} U_k(\phi,h^*,h)|_{h^*=h=0}$ and $Z_k^{(10)}(\phi)\equiv\partial_{h^*} Z_k(\phi,h^*,h)|_{h^*=h=0}$ with initial conditions $U_{\kin}^{(10)}(\phi)=-e^{i\sqrt{8\pi}a\phi}$ and $Z_{\kin}^{(10)}(\phi)=0$~\cite{not2}. 

We have computed the expectation value of the exponential fields $e^{in\sqrt{2\pi K}\varphi}$ ($n$ integer). These are the natural fields to consider in the sine-Gordon model. For instance, in one-dimensional quantum fluids, they arise from products of single-particle fields. The FRG results for $1\leq n\leq 5$ are shown in Fig.~\ref{fig_conjecture}. 

For $n=1$ we find an excellent agreement between the FRG results and the conjecture, with a difference $\epsabs$ well below 0.01 for all values of $K$. The relative disagreement $\epsrel$ is small for $K<0.5$ but increases for larger values of $K$ and becomes of order of 100\% for $K$ near 1. For these values of $K$ however, the expectation value $\mean{e^{i\sqrt{2\pi K}\varphi}}$ is very small and what matters is $\epsabs$. 

Note that the Lukyanov-Zamolodchikov conjecture breaks down in the vicinity of $K=1/n$ since the expectation value $\mean{e^{in\sqrt{2\pi K}\varphi}}$ given by Eq.~(\ref{conjecture}) diverges when $K\to 1/n$~\cite{not8}. This explains the steep upturn near $K=1/n$ of the lines showing the conjecture in Fig.~\ref{fig_conjecture}. Decreasing the value of $u/\Lamb^2$ confines the upturn more and more to the vicinity of $1/n$. 

For $n\geq 2$, $\epsrel$ behaves similarly to the case $n=1$ but $\epsabs$ is also a monotonously increasing function of $K$ (see the inset in Fig.~\ref{fig_conjecture}). $\epsabs$ remains nevertheless below 0.01 up to values of $K$ very close to $1/n$; for $u/\Lamb^2=10^{-4}$ this is the case for $K=0.49$ (and $n=2$), $K=0.33$ ($n=3$) and $K=0.248$ ($n=4$). Moreover $\epsrel$ decreases when $u/\Lamb^2$ is reduced (which extends the domain of validity of the conjecture to higher values of $K$, i.e. to values of $K$ closer to $1/n$). For instance, for $n=2$ and $K=0.49$, we find $\epsrel=78/72/66$\% (while $\epsabs=0.097/0.0097/0.00098$) for $u/\Lamb^2=10^{-3}/10^{-4}/10^{-5}$. We therefore ascribe the apparent disagreement between the FRG results and the conjecture near $K=1/n$ to the breakdown of the latter when $K\to 1/n$. In fact the change of concavity in the curves showing $\epsabs$ in the inset of Fig.~\ref{fig_conjecture} suggests that the conjecture might deviate from the correct result well before $K=1/n$ (e.g. $K\sim 0.4$ for $n=2$ and $u/\Lamb^2=10^{-4}$). A conservative estimate is that the FRG reproduces Eq.~(\ref{conjecture}), in the domain of validity of the conjecture, to an accuracy (defined as the minimum of $\epsabs$ and $\epsrel$) better than 0.01.

\paragraph{Conclusion.}

Contrary to the perturbative RG~\cite{not6,Wiegmann78,Amit80}, which correctly predicts the phase diagram of the quantum sine-Gordon model but fails to describe the massive phase, the nonperturbative FRG allows us to continue the flow into the strong-coupling regime and thus compute the low-energy properties of the massive phase. The fact that FRG captures genuinely nonperturbative topological excitations, namely (anti)solitons and breathers, proves its efficiency and is reminiscent of its ability to describe most universal properties of the BKT transition in the linear O(2) model~\cite{Graeter95,Gersdorff01,Jakubczyk14} 
for which it is widely admitted that topological defects (vortices) play a crucial role.

The FRG result for the expectation value $\mean{e^{in\sqrt{2\pi K}\varphi}}$ of the exponential fields is in very good agreement with the conjecture proposed by Lukyanov and Zamolodchikov~\cite{Lukyanov97}. The minimum of the relative and absolute disagreements is smaller than 0.01 for all values of $n$ except in the immediate vicinity of $K=1/n$ where the conjecture breaks down. This undoubtedly provides us with a very strong support of the Lukyanov-Zamolodchikov conjecture. We also stress that FRG allows one to obtain $\mean{e^{in\sqrt{2\pi K}\varphi}}$ for all values of $K$ whereas the conjecture is limited to the range $K<1/n$.   

Finally we would like to point out that the nonperturbative FRG approach presented in this Letter opens up the possibility to study various nonintegrable extensions of the quantum sine-Gordon model where both perturbative RG and exacts methods are inoperative in the strong-coupling phase.  

\paragraph{Acknowledgment.} We thank P. Azaria for enlightening discussions and a critical reading of the manuscript.


%

\setcounter{equation}{0}
\newpage

\def\kin{k_{\rm in}} 
\def\Msol{M_{\rm sol}} 
\def\Mcl{M_{\rm cl}} 

\begin{center}
{\bf \Large Supplemental Material}
\end{center}

We consider the sine-Gordon model defined by the Euclidean action 
\beq
S[\varphi] = \int d^2r \llbrace \half (\nablabf\varphi)^2 - u \cos(\beta\varphi) \rrbrace , 
\label{action} 
\eeq
where regularization with a UV momentum cutoff $\Lambda$ is implied and $u/\Lamb^2,\beta>0$ are dimensionless parameters. In the following, instead of the parameter $\beta$, we will often use the Luttinger parameter $K=\beta^2/8\pi$. 

\section{I. Exact flow equation}

To implement the functional renormalization-group (FRG) approach we add to the action the infrared regulator term
\beq
\Delta S_k[\varphi] = \half \sum_\q \varphi(-\q) R_k(\q) \varphi(\q) 
\label{DeltaS}
\eeq 
such that fluctuations are smoothly taken into account as $k$ is lowered from the microscopic scale $\kin$ down to 0~\cite{Berges02sup,Delamotte12sup,Kopietz_booksup}. The regulator function in~(\ref{DeltaS}) is defined by 
\beq
R_k(\q) = Z_k \q^2 r\left( \frac{\q^2}{k^2} \right) ,
\eeq 
where the function $r(y)$ satisfies $r(0)=\infty$ and $r(\infty)=0$, and $Z_k$ is a field renormalization factor defined below. Thus $\Delta S_k$ suppresses fluctuations with momenta $|\q|\ll k$ but leaves unaffected those with $|\q|\gg k$. Various choices for $r(y)$ are discussed below. 

The partition function 
\beq
\calZ_k[J] = \int \calD[\varphi]\, e^{-S[\varphi] - \Delta S_k[\varphi] +\int d^2r\,J\varphi} 
\eeq
is $k$ dependent. The scale-dependent effective action 
\beq
\Gamma_k[\phi] = - \ln  \calZ_k[J] + \int d^2r\, J\phi - \Delta S_k[\phi] 
\eeq
is defined as a modified Legendre transform of $-\ln\calZ_k[J]$ which includes the subtraction of $\Delta S_k[\phi]$. Here $\phi(\r)=\mean{\varphi(\r)}$ is the order parameter (in the presence of the external source $J$). Provided that the initial value $\kin$ of $k$ is sufficiently large wrt the UV momentum cutoff $\Lamb$ of the sine-Gordon model, all fluctuations are completely frozen by $\Delta S_{\kin}[\varphi]$ and $\Gamma_{\kin}[\phi]=S[\phi]$. On the other hand the effective action of the sine-Gordon model is given by $\Gamma_{k=0}$ since $\Delta S_{k=0}=0$. The FRG approach aims at determining $\Gamma_{k=0}$ from $\Gamma_\Lamb$ using Wetterich's equation~\cite{Wetterich93sup,Ellwanger94sup,Morris94sup} 
\beq
\dk \Gamma_k[\phi] = \half \Tr \llbrace \dk R_k \bigl(\Gamma^{(2)}_k[\phi]+R_k \bigr)^{-1} \rrbrace ,
\label{rgeq}
\eeq 
where $\Gamma^{(2)}_k[\phi]$ denotes the second-order functional derivative of $\Gamma_k[\phi]$. 

\section{II. Derivative expansion}

To solve the exact flow equation~(\ref{rgeq}) we use a derivative expansion of the scale-dependent effective action. Such an expansion is made possible by the regulator term $\Delta S_k$, which ensures that all vertices $\Gamma^{(n)}_k(\q_1\cdots\q_n)$ are smooth functions of momenta $\q_i$ and can be expanded in powers of $\q^2_i/\max(k,M)^2$ when $|\q_i|\ll \max(k,M)$ with $M$ the smallest mass in the spectrum. Thus the derivative expansion is sufficient to compute the soliton and breather masses in the massive phase as well as the expectation value of the exponential fields.  

To second order of the derivative expansion, the effective action 
\beq
\Gamma_k[\phi] = \int d^2r \biggl\lbrace \half Z_k(\phi) (\nablabf\phi)^2 + U_k(\phi) \biggr\rbrace 
\label{GamDE} 
\eeq 
is fully determined by two functions of the field, $Z_k(\phi)$ and $U_k(\phi)$, which are periodic on the interval $]-\pi/\beta,\pi/\beta]$. In practice we consider the dimensionless functions 
\beq
\tilde U_k(\phi) = \frac{U_k(\phi)}{Z_k k^2} , \quad \tilde Z_k(\phi) = \frac{Z_k(\phi)}{Z_k}, 
\eeq
where $Z_k$ is defined from the condition 
\beq
\frac{\beta}{2\pi} \int_{-\pi/\beta}^{\pi/\beta} d\phi\, \tilde Z_k(\phi) = 1 .  
\label{Zdef}
\eeq
The flow equations for the functions $\tilde U_k(\phi)$ and $\tilde Z_k(\phi)$, obtained by inserting~(\ref{GamDE}) into~(\ref{rgeq}), are given in Appendix~\ref{app_rgeq}.

\section{III. Choice of the regulator function $R_k$} 

\begin{figure} 
\includegraphics[width=7cm]{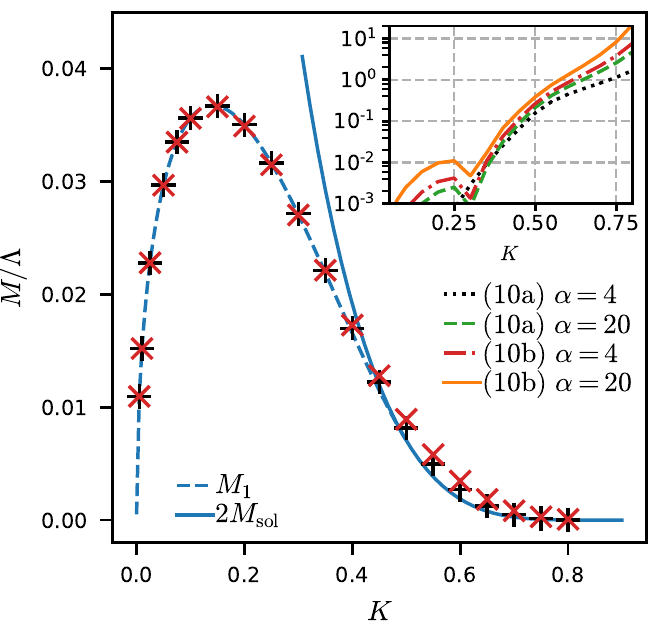}
\caption{Mass $M$ as obtained from the FRG approach with the regulator functions~(\ref{ra}) ($\textcolor{black}{+}$) and (\ref{rb}) ($\textcolor{red}{\times}$) and $\alpha=4$. The solid and dashed lines show the exact values of $2\Msol$ and $M_1$ (the latter being defined only for $K\leq 1/2$). The inset shows the relative error of the FRG result for the same regulator functions and $\alpha=4$ or $\alpha=20$. The regulator function~(\ref{rc}) gives results (not shown) similar to (\ref{ra}). (Here and in the following figures: $\Lamb=1$ and $u/\Lamb^2=10^{-3}$.)}
\label{fig_Rk} 
\end{figure}

The initial value of the running momentum scale must satisfy $\kin\gg\Lamb$ in order to ensure that all fluctuations are frozen and $\Gamma_{\kin}[\phi]=S[\phi]$. In practice we take $\kin\sim 100\Lamb$. Figure~\ref{fig_Rk} shows the mass $M$ obtained from FRG with different regulator functions: 
\begin{subequations}
\begin{empheq}[left={r(y)=}\empheqlbrace]{align}
& \displaystyle \frac{\alpha}{e^y-1} \label{ra} \\ 
& \displaystyle \alpha \frac{(1-y)^2}{y} \Theta(1-y) \label{rb}\\
& \displaystyle \alpha \frac{e^{-y}}{y}(1+ \gamma y) \label{rc}
\end{empheq} 
\label{Rk}  
\end{subequations}
($\alpha,\gamma$ are free parameters). In the range $K\leq 0.4$, the accuracy of the FRG result is better than 2\% regardless of the choice of the regulator. For $K>0.4$ it deteriorates and the results become strongly dependent on the regulator. Note however that although the relative error can be larger than 100\%, the absolute error remains small since the mass $M$ goes to zero as $K\to 1$. 

\section{IV. The nonuniversal scale factor $b$} 

In a theory with a UV momentum cutoff $\Lamb$ the soliton mass is given by~\cite{Zamolodchikov95sup}
\beq
\Msol = \frac{2 \Gamma\left( \frac{K}{2-2K} \right)}{ \sqrt{\pi}\Gamma\left( \frac{1}{2-2K} \right) }
\left[ \frac{ \Gamma(1-K) }{ \Gamma(K) } \frac{\pi}{2} u \Lamb_{\calR}^{-2K} \right]^{\frac{1}{2-2K}} .
\label{Msol}
\eeq
The value of $\Lamb_{\calR}$ (which depends on $\Lamb$) can be determined by considering the one-loop correction to the mass $M_{1,\rm cl}=\sqrt{8\pi Ku}$ of the lightest breather in the semiclassical limit $K\to 0$~\cite{Zamolodchikov95sup}, i.e. 
\beq
\frac{M_1^2}{M_{1,\rm cl}^2} = 1 + 2K \ln \left( \frac{M_{1,\rm cl}e^C}{2\Lamb_{\calR}} \right) ,
\label{fdef1}
\eeq
where $C$ is the Euler constant. An elementary calculation based on the action~(\ref{action}) gives~\cite{[{Recall that the mass $M_1$ of the lightest breather can be simply obtained by considering fluctuations of the $\varphi$ field about $\varphi=0$. See, e.g., }]Rajaraman_book}    
\beq
\frac{M_1^2}{M_{1,\rm cl}^2} 
= 1 - 2K \int_0^\infty dq \frac{q}{q^2+ M_{1,\rm cl}^2} f_\Lamb(q) , 
\label{fdef} 
\eeq
where $f_\Lamb(q)$ is a momentum cutoff function (e.g. $f_\Lamb(q)=\Theta(\Lamb-q)$ for a hard cutoff). By comparing~(\ref{fdef1}) and (\ref{fdef}), we find that $\Lamb$ and $\Lamb_{\calR}$ are related by 
\beq
\ln \left( \frac{2\Lamb_{\calR}e^{-C}}{M_{1,\rm cl}} \right) = \int_0^\infty dq \frac{q}{q^2+M_{1,\rm cl}^2} f_\Lamb(q) .
\label{bdef} 
\eeq 
For $M_{1,\rm cl}\ll\Lamb$, the rhs gives $\ln(\Lamb/M_{1,\rm cl})+\const$ and Eq.~(\ref{bdef}) allows us to determine the scale factor $b=\Lamb_{\calR}/\Lamb$ for a given cutoff function $f_\Lamb(q)$ (for a hard cutoff $b=e^C/2$). The soliton mass in the model defined with a momentum cutoff $\Lamb$ can therefore be written as 
\beq
\Msol = b\Lamb \frac{2 \Gamma\left( \frac{K}{2-2K} \right)}{ \sqrt{\pi}\Gamma\left( \frac{1}{2-2K} \right) }
 \left[ \frac{ \Gamma(1-K) }{ \Gamma(K) } \frac{\pi u}{2(b\Lamb)^2} \right]^{\frac{1}{2-2K}} ,
\label{Msol2} 
\eeq 
where $b=\Lamb_{\calR}/\Lamb$ is obtained from~(\ref{bdef}).

\section{V. Universality} 

\begin{figure} 
\includegraphics[width=7cm]{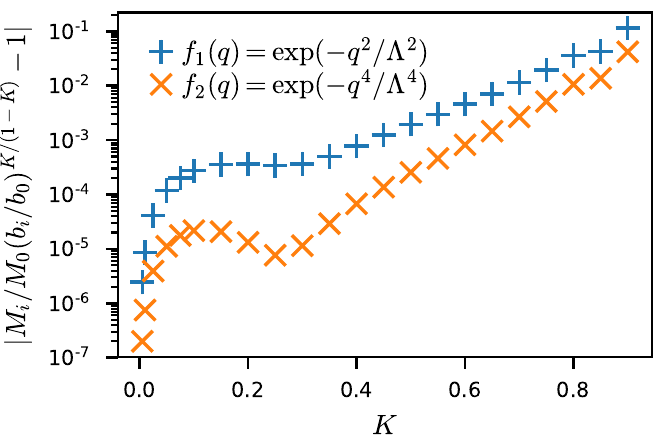}
\caption{$|(M_i/M_0)(b_i/b_0)^{K/(1-K)}-1|$ for fixed values of $\Lamb$ and $u/\Lamb^2$ and two different cutoff functions, $f_1(q)$ and $f_2(q)$. $M_0$ is the mass obtained with the hard cutoff $f_\Lamb(q)=\Theta(\Lamb-q)$. The scale parameter $b=\Lamb_{\calR}/\Lamb$ is obtained from~(\ref{bdef}).}
\label{fig_universality} 
\end{figure}

Equation~(\ref{Msol2}) implies universality in the sense that $\Msol/\Lamb$ is a universal function of $K$ and $u/\Lamb^2$ up to the scale factor $b$. This universality is due to the flow being controlled by the Gaussian fixed point $u=0$ for $\Msol\ll k\ll\Lamb$. In this momentum range all correlation functions exhibit the same scaling, e.g. $\mean{e^{i\beta\varphi(\r)} e^{-i\beta\varphi(\r')}} \sim (|\r-\r'|\Lamb)^{-4K}$, up to a nonuniversal prefactor that depends on the cutoff function $f_\Lamb(q)$. Various theories, corresponding to different implementations of the cutoff (or, equivalently, different normalizations of the correlation functions), are simply related by the scale factor $b$.
 
In Fig.~\ref{fig_universality} we show the ratio 
\beq
\frac{M_i}{M_0} \left(\frac{b_i}{b_0}\right)^\frac{K}{1-K} , 
\label{Mratio}
\eeq
where $M_0$ is the mass obtained with the hard cutoff $f_\Lamb(q)=\Theta(\Lamb-q)$ and $M_i$ that obtained with the cutoff $f_1(q)=e^{-q^2/\Lamb^2}$ or $f_2(q)=e^{-q^4/\Lamb^4}$. The scale parameter $b=\Lamb_{\calR}/\Lamb$ is obtained from~(\ref{bdef}).
Universality implies that the ratio~(\ref{Mratio}) is equal to one. This property is satisfied with a very high accuracy (the ratio differs from one by less than $10^{-3}$) in the range $K\leq 0.4$ where the mass $M$ is obtained with high precision. Not surprisingly, deviations from universality are more pronounced for $K>0.4$ due to the lesser accuracy of the FRG results.

\section{VI. Semiclassical limit} 

\begin{figure} 
\includegraphics[width=7cm]{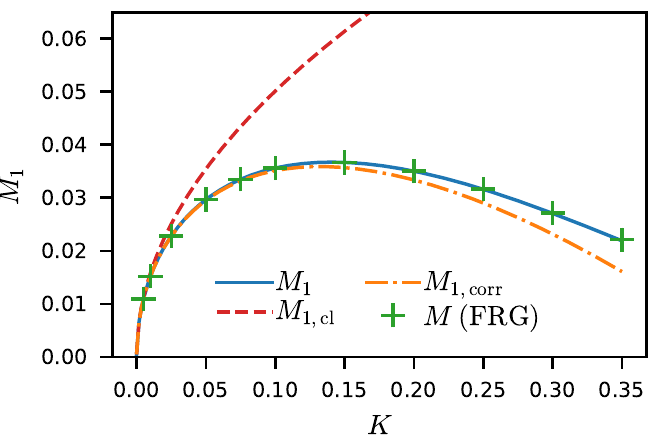}
\caption{Mass $M_1$ of the lightest breather for $K\ll 1$. The FRG result is compared to the leading-order (semiclassical) value $M_{1,\rm cl}=\sqrt{8\pi Ku}$ and the next-to-leading-order value $M_{1,\rm corr}$ given by~(\ref{fdef1}) as well as the exact result $M_1=2\Msol \sin\bigl(\frac{\pi K}{2(1-K)}\bigr)$ given by~(\ref{Msol2}).}
\label{fig_semiclassical} 
\end{figure}

In the limit $K\ll 1$, the leading correction to the mass $M_{1,\rm cl}$ of the lightest breather is given by Eq.~(\ref{fdef1}) with $\Lamb_{\calR}=b\Lamb$. When $K\to 0$ the partition function can be computed by a saddle-point approximation and the initial value $\Gamma_{\kin}[\phi]=S[\phi]$ of the effective action gives the exact result $M_{1,\rm cl}=\sqrt{8\pi Ku}$ (the flow does not bring any correction to $\Gamma_{\kin}[\phi]=S[\phi]$ when $K\to 0$). In Fig.~\ref{fig_semiclassical} we compare the FRG result for $M_1$ with the exact value, the semiclassical limit $K\to 0$ and the perturbative result~(\ref{fdef1}). For $K\lesssim 0.1$, the agreement between FRG and the exact solution is nearly perfect.

\section{VII. Expectation value of exponential fields} 

To compute the expectation value of the exponential fields, we add to the action the source term 
\beq
S_h[\varphi] = - \int d^2r (h^* e^{in\sqrt{2\pi K}\varphi} + \cc )  
\eeq
so that $\mean{e^{in\sqrt{2\pi K}\varphi(\r)}}$ can be obtained from $\ln\calZ_k[J,h^*,h]$ by functional differentiation wrt $h^*(\r)$. 
The scale-dependent effective action reads 
\beq
\Gamma_k[\phi,h^*,h] = -\ln \calZ_k[J,h^*,h] + \int d^2r\, J\phi - \Delta S_k[\phi]  
\eeq 
and satisfies the flow equation
\beq
\dt \Gamma_k[\phi,h^*,h] = \half \Tr \Bigl\lbrace \dt R_k \bigl(\Gamma^{(2)}[\phi,h^*,h]+R_k\bigr)^{-1} \Bigr\rbrace 
\eeq
with initial conditions $\Gamma_{\kin}[\phi,h^*,h] = S[\phi] + S_h[\phi]$. 

\begin{figure}[t]
\centerline{\includegraphics{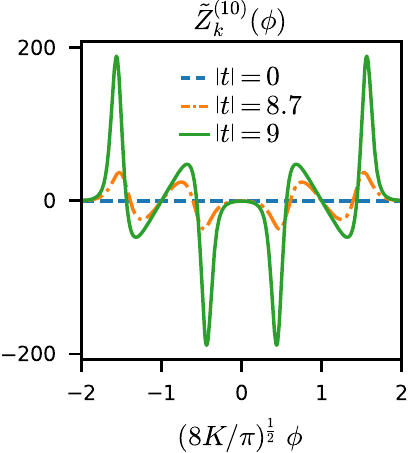}
\includegraphics{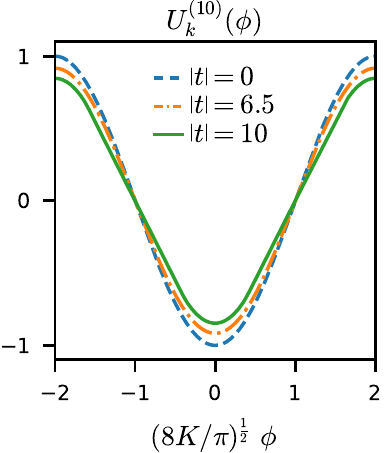}}
\caption{$\tilde Z^{(10)}_k(\phi)$ and $U^{(10)}_k(\phi)$ for various values of $t=\ln(k/\kin)$ and $n=1$. 
} 
\label{fig_ZU10} 
\end{figure}

We consider the ansatz 
\begin{align}
\Gamma_k[\phi,h^*,h] ={}& \int d^2r \biggl\lbrace \half Z_k(\phi,h^*,h) (\nablabf\phi)^2 \nonumber \\ 
& + U_k(\phi,h^*,h) \biggr\rbrace .
\end{align} 
The expectation value of interest reads 
\beq
\mean{e^{in\sqrt{2\pi K}\varphi(\r)}} = - \frac{\partial U_k(\phi,h^*,h)}{\partial h^*}\biggl|_{h^*=h=0 \atop \phi=0} \equiv - U_k^{(10)}(\phi=0) ,
\eeq
noting that $\phi=0$ corresponds to the minimum of the potential $U(\phi,h^*,h)$ when $h^*=h=0$. The RG equation for $U_k^{(10)}(\phi)$ involves $Z_k^{(10)}(\phi)\equiv \partial_{h^*}Z_k(\phi,h^*,h)|_{h^*=h=0}$. In practice we consider the dimensionless functions 
\beq
\tilde U_k^{(10)}(\phi) = \frac{U_k^{(10)}(\phi)}{Z_k k^2}, \quad \tilde Z_k^{(10)}(\phi) = \frac{Z_k^{(10)}(\phi)}{Z_k} .
\eeq
The flow equations satisfied by $\tilde U_k^{(10)}(\phi)$ and $\tilde Z_k^{(10)}(\phi)$ are given in Appendix~\ref{app_rgeq}. The results obtained in the massive phase are shown in Fig.~\ref{fig_ZU10} for $n=1$.

\newpage

\begin{widetext} 
\section{Appendix A: Flow equations} 
\label{app_rgeq} 

The flow equations read  
\begin{align} 
\dt \tilde U_k ={}& (\eta_k-2) \tilde U_k + \frac{1}{4\pi Z_k} l_{0,0}^2 , \\ 
\dt \tilde Z_k ={}& \eta_k \tilde Z_k -\frac{1}{4\pi Z_k} \Bigl[  2 \tilde U_k''' l_{0,2}^6 \tilde Z_k'-2 \tilde U_k''' l_{2,0}^2 \tilde Z_k'+
   \tilde U_k'''{}^2 l_{0,2}^4+ l_{1,0}^2 \tilde Z_k''+ l_{0,2}^8
   \tilde Z_k'{}^2- \frac{5}{2} l_{2,0}^4 \tilde Z_k'{}^2 \Bigr] 
\end{align}
and 
\begin{align}  
\dt \tilde U^{(10)}_k ={}& (\eta_k-2) \tilde U^{(10)}_k 
- \frac{1}{4\pi Z_k} \Bigl[ l_{1,0}^2 \tilde U^{(10)}_k{}'' + l_{1,0}^4 \tilde Z^{(10)}_k \Bigr] , \\ 
\dt \tilde Z^{(10)}_k ={}& \eta_k \tilde Z^{(10)}_k -\frac{1}{4\pi Z_k} \biggl\{ 2 l_{0,2}^6 \left[\tilde U_k^{(10)}{}''' \tilde Z_k'+\tilde U_k''' \tilde Z_k^{(10)}{}'\right]+2
   \tilde U_k^{(10)}{}''' \tilde U_k''' l_{0,2}^4-8 \tilde U_k''' l_{1,2}^6 \tilde U_k^{(10)}{}''
   \tilde Z_k'
   \nonumber \\ & 
   +4 \tilde U_k''' l_{3,0}^2 \tilde U_k^{(10)}{}'' \tilde Z_k'
   -4 \tilde U_k'''{}^2 l_{1,2}^4 \tilde U_k^{(10)}{}''-4 l_{1,2}^8 \tilde U_k^{(10)}{}''
   \tilde Z_k'{}^2+5 l_{3,0}^4 \tilde U_k^{(10)}{}'' \tilde Z_k'{}^2-l_{2,0}^2
   \Bigl[ 2 \tilde U_k^{(10)}{}''' \tilde Z_k'
    \nonumber \\ &
   +\tilde U_k^{(10)}{}'' \tilde Z_k''+2 \tilde U_k'''
   \tilde Z_k^{(10)}{}'\Bigr] -8 \tilde U_k''' \tilde Z_k^{(10)}{} l_{1,2}^8 \tilde Z_k'+2 \tilde U_k'''
   \tilde Z_k^{(10)}{} l_{2,1}^6 \tilde Z_k'+8 \tilde U_k''' \tilde Z_k^{(10)}{} l_{3,0}^4 \tilde Z_k'-4
   \tilde U_k'''{}^2 \tilde Z_k^{(10)}{} l_{1,2}^6 
   \nonumber \\ &
   +\tilde U_k'''{}^2
   \tilde Z_k^{(10)}{} l_{2,1}^4+\tilde U_k'''{}^2 \tilde Z_k^{(10)}{} l_{3,0}^2+l_{1,0}^2
   \tilde Z_k^{(10)}{}''+2 l_{0,2}^8 \tilde Z_k^{(10)}{}' \tilde Z_k'-l_{2,0}^4 \Bigl[ 5 \tilde Z_k^{(10)}{}'
   \tilde Z_k'+\tilde Z_k^{(10)}{} \tilde Z_k''\Bigr] 
   \nonumber \\ &
   -4 \tilde Z_k^{(10)}{} l_{1,2}^{10}
   \tilde Z_k'{}^2+\tilde Z_k^{(10)}{} l_{2,1}^8 \tilde Z_k'{}^2+8 \tilde Z_k^{(10)}{}
   l_{3,0}^6 \tilde Z_k'{}^2 \biggr\}  ,
\end{align}
where $t=\ln(k/\kin)$ and $\eta_k=-\dt\ln Z_k$. The equation for $\eta_k$ is simply derived from~(\ref{Zdef}). To alleviate the notations we do not write explicitly the $\phi$ dependence of $\tilde U_k(\phi)$, $\tilde Z_k(\phi)$, $\tilde U^{(10)}_k(\phi)$ and $\tilde Z^{(10)}_k(\phi)$. The initial conditions are 
\beq
\tilde U_{\kin}(\phi) = - (u/\kin^2) \cos(\beta\phi), \quad
\tilde Z_{\kin}(\phi) = 1 , \quad
\tilde U^{(10)}_{\kin}(\phi) = - e^{in\sqrt{2\pi K}\phi}/\kin^2 , \quad \tilde  Z^{(10)}_{\kin}(\phi)=0 
\eeq
and $Z_{\kin}=1$. The threshold function $l^d_{n,m}$ is defined in Appendix~\ref{app_threshold}.

\subsection*{Weak-coupling limit} 

The perturbative flow equations are recovered by retaining a single harmonic of the potential, i.e. $\tilde U_k(\phi)=-\tilde u_{k} \cos(\beta\phi)$, and neglecting the flow of $\tilde Z_k(\phi)$. Using 
\beq 
l^2_{0,0}(\tilde U'',1,0) = l^2_{0,0}(0,1,0) - l^2_{1,0}(0,1,0) \tilde U'' +\calO(\tilde U''{}^2) , 
\eeq
where $l^2_{1,0}(0,1,0)=1$ is universal (i.e. independent of the regulator function $r$), one obtains 
\beq
\begin{split} 
\dt \tilde u_{k} &= -2\tilde u_k (1-K_k) + \calO(\tilde u_k^2) , \\ 
\dt K_k &= \tilde u_k^2  (8\pi K^2)^2 l_{0,2}^4(0,1,0) + \calO(\tilde u_k^3) ,
\end{split}
\label{rgpert}
\eeq
where $K_k=K/Z_k$ is a ``running'' Luttinger parameter. Sufficiently far away from the BKT point $(K_k=1,\tilde u_k=0)$, one can neglect the renormalization of $K_k$ when $\tilde u_k\to 0$ and the flow equations become $\dt \tilde u_k=-2\tilde u_k(1-K)$ to leading order. For $K<1$, $\tilde u_k$ grows as $k^{2K-2}$. This is the massive phase studied in the Letter. When $1-K_k$ and $\tilde u_k$ are of the same order, both equations in~(\ref{rgpert}) must be solved simultaneously and one recovers the scaling behavior of the BKT phase transition~\cite{[{See, e.g., }]Chaikin_booksup}.

\section{Appendix B: Threshold functions}
\label{app_threshold}

The threshold function $l^d_{n_1,n_2}\equiv l^d_{n_1,n_2}(\tilde U_k'',\tilde Z_k,\eta_k)$ is defined by 
\beq
\begin{split}
l^d_{0,0} &= \int_0^\infty d\tilde q\, \tilde q^{d-1} \frac{\dot R_k}{Z_k k^2} \tilde G , \\ 
l^d_{n_1,n_2} &= - \tdt \int_0^\infty d\tilde q\, \tilde q^{d-1} \tilde G^{n_1}  \tilde G'{}^{n_2}  \quad \mbox{if} \quad (n_1,n_2)\neq (0,0),
\end{split}
\eeq 
where 
\beq
\begin{split} 
\tilde G &= [(\tilde Z_k+r)y + \tilde U_k'']^{-1} , \qquad 
\tilde \dt \tilde G = - \frac{\dot R_k}{Z_k k^2} \tilde G^2 , \\ 
\tilde G' &= - \tilde G^2 [\tilde Z_k + r + yr' ] , \qquad
\tilde \dt G' = 2 \frac{\dot R_k}{Z_k k^2} \tilde G^3 [\tilde Z_k + r + yr' ] - \frac{\dot R_k'}{Z_k} \tilde G^2 , 
\end{split}
\eeq
and 
\beq 
\begin{split} 
R_k &= Z_k k^2 y r, \qquad
\dot R_k = - Z_k k^2 y (\eta_k r+2yr') , \\
R_k' &= Z_k (r+yr') , \qquad 
\dot R_k' = - Z_k [ \eta_k(r+yr') +2 y (2r'+yr'')] , 
\end{split}
\eeq
with $y=\tilde q^2=\q^2/k^2$, $r\equiv r(y)$, $r'\equiv r'(y)$, etc. The prime denotes a derivative wrt $y$, the dot a derivative wrt $t$  and $\tdt$ acts only on the $k$ dependence of $R_k$. 

\end{widetext}


\begin{thebibliography}{42}
\bibitem [{not({\natexlab{a}})}]{not5}%
  \BibitemOpen
  \href@noop {} {} \bibinfo {note} {Classically the
  topological charge of a static field configuration is defined by
  $Q=(\beta/2\pi) \int_{-\infty}^\infty dx\, \partial_x
  \varphi=(\beta/2\pi)[\varphi(\infty)-\varphi(-\infty)]$. In the quantum
  theory $Q$ becomes a topological quantum
  number~\cite{Rajaraman_book}.}\BibitemShut {Stop}%
\bibitem [{not({\natexlab{b}})}]{not4}%
  \BibitemOpen
  \href@noop {} {} \bibinfo {note} {For $a=\sqrt{K}$ the
  expectation value $\mean{e^{i\sqrt{8\pi}a\varphi}}$ is simply related to the
  free energy which is exactly known. For $K=1/2$ the exact value is deduced
  from the mapping onto the massless Thirring
  model~\cite{Lukyanov97}.}\BibitemShut {Stop}%
\bibitem [{not({\natexlab{c}})}]{not6}%
  \BibitemOpen
  \href@noop {} {} \bibinfo {note} {The perturbative
  renormalization group is a popular method to study the sine-Gordon
  model~\cite{Gogolin_book,Tsvelik_book,Giamarchi_book}. In the massive phase
  however, perturbative scaling equations lead to an unphysical divergence of
  the coupling constant $u$. The phase diagram can be determined but the
  understanding of the physical properties of the massive phase is out of reach
  of the perturbative approach.}\BibitemShut {Stop}%
\bibitem [{not({\natexlab{e}})}]{not7}%
  \BibitemOpen
  \href@noop {} {} \bibinfo {note} {Note that we do not
  rescale the field in order to preserve the $2\pi/\beta$ periodicity of
  $Z_k(\phi)$ and $U_k(\phi)$.}\BibitemShut {Stop}%
\bibitem [{not({\natexlab{f}})}]{not8}%
  \BibitemOpen
  \href@noop {} {} \bibinfo {note} {Near $K=1$ and $u=0$ the
  behavior of the system is controlled by the BKT fixed point, which explains
  why the conjecture cannot be valid arbitrary close to $K=1$.}\BibitemShut
  {Stop}%
\end{thebibliography}

%

\end{document}